\documentclass[journal,twoside,10pt]{IEEEtran}
\usepackage{cite}
\usepackage{amsmath,amssymb,amsfonts}

\usepackage[caption=false,font=footnotesize]{subfig}
\usepackage{graphicx}
\usepackage{textcomp}
\usepackage{xcolor}

\usepackage{booktabs}
\usepackage{algorithm}
\usepackage{algpseudocode}

\def\BibTeX{{\rm B\kern-.05em{\sc i\kern-.025em b}\kern-.08em
    T\kern-.1667em\lower.7ex\hbox{E}\kern-.125emX}}
\newtheorem{remark}{Remark}

\begin{document}
\title{Covariance-Guided DFT Beam Selection for Beamspace ESPRIT in Hybrid mmWave Sensor Arrays}
\author{R{\i}fat Volkan \c{S}enyuva
\thanks{\copyright{} 2026 IEEE. Personal use of this material is permitted.
Permission from IEEE must be obtained for all other uses, in any current or
future media, including reprinting/republishing this material for advertising
or promotional purposes, creating new collective works, for resale or
redistribution to servers or lists, or reuse of any copyrighted component of
this work in other works. Accepted for publication in \textit{IEEE Sensors
Journal}. DOI to be assigned.}%
\thanks{R.~V.~\c{S}enyuva is with the Dept.\ of Electrical--Electronics
Engineering, Maltepe University, 34857 Istanbul, Turkey
(e-mail: rifatvolkansenyuva@maltepe.edu.tr).}}

\maketitle

\begin{abstract}
Accurate direction-of-arrival estimation with hybrid analog--digital millimeter-wave sensor arrays is important for localization, environment sensing, and measurement beam control for sensing applications. However, the limited number of radio-frequency chains and training beams in practical hardware makes it difficult to approach the angular resolution of fully digital arrays. This paper develops a covariance-guided discrete Fourier transform (DFT) beam selection framework tailored to beamspace ESPRIT for hybrid millimeter-wave receivers. A short hybrid training phase realizes a virtual centro-symmetric subarray and yields a sample covariance that is processed by forward--backward averaging, nonnegative least-squares power and noise fitting, and a Toeplitz positive-semidefinite projection to reconstruct a denoised full-aperture covariance matrix. This covariance is then used to score and select, within each coarse sector, small contiguous blocks of DFT beams that concentrate signal energy and preserve effective aperture under a strict beam budget. The selected beams feed a sparse beamspace ESPRIT stage that operates only on actually available adjacent beam pairs, so that the overall complexity is dominated by a single low-dimensional ESPRIT call. Monte Carlo simulations for a thirty-two-element uniform linear array with three paths indicate that, in the considered scenarios, the proposed method can reduce the gap to the Cram\'er--Rao bound, lower the failure rate, and provide favorable accuracy--runtime trade-offs compared with a sectorization-based baseline built from the same codebook and estimator.
For the unitary DFT codebook studied here, the fine-stage beam selector reduces to a covariance-guided contiguous energy-window rule; the broader score formulation also accommodates non-unitary effective beam dictionaries arising from hardware non-idealities.
\end{abstract}


\section{Introduction}
\label{sec:intro}
\IEEEPARstart{H}{ybrid} analog/digital (HAD) arrays and DFT beamspace processing are key enablers of millimeter-wave (mmWave) sensing and communication, since they reduce radio-frequency (RF) hardware cost while preserving a large effective aperture. In many sensing applications, such as mmWave radar, localization, and environment mapping, the front-end must estimate directions-of-arrival (DoAs) or virtual angles accurately over a practical range of array signal-to-noise ratios (ASNRs), numbers of snapshots, and calibration errors, while operating with only a few RF chains and a small set of pre-defined DFT beams. These constraints motivate DoA estimation algorithms that exploit the covariance structure of the received data and the geometry of mmWave sensor arrays, rather than relying on fully digital architectures.

Classical subspace-based DoA estimation methods, such as MUSIC and ESPRIT, achieve high resolution when a fully digital array is available and the sample covariance matrix is well conditioned. Beamspace processing further reduces dimensionality by projecting the array outputs onto fixed DFT beams, while preserving most of the information in the dominant angular region. More recently, Tensor-ESPRIT algorithms in DFT beamspace have been used for gridless channel and DoA estimation in hybrid mmWave MIMO systems, leveraging the structure of uniform linear and planar arrays and the Kronecker separability of the array manifold \cite{TensorESPRIT_Beamspace}. Building on this line of work, tensor-based ESPRIT frameworks for joint channel and target parameter estimation in massive MIMO-ISAC systems have been proposed that exploit multi-dimensional Vandermonde structure across angle, delay, and Doppler dimensions using a large beam budget and a shared training pattern \cite{ZhangCheng2024TWC}. The present work addresses a structurally different regime: a dedicated sensing receiver with a strict per-sector DFT beam budget ($K_g{=}2$ beams per sector), where multi-dimensional tensor exploitation is precluded by hardware constraints and the estimator reduces to a single low-dimensional 1-D ESPRIT call on a small contiguous beam subset. Complementing these works, the analysis in \cite{Asilomar_ESPRIT_MSE} derives closed-form mean-squared error (MSE) expressions for 1-D ESPRIT in DFT beamspace in terms of physical parameters, and \cite{ApertureReduction_Beamspace} shows that mixing a few adjacent DFT beams is equivalent to reducing the aperture in element space. These results highlight that, under DFT beamspace processing, both the effective aperture and the choice of adjacent beams strongly influence the DoA estimation accuracy.

In related work, sparse/gridless DoA estimators (e.g., sparse Bayesian learning and atomic norm methods) can deliver high resolution but often require iterative optimization and typically assume access to a high-quality full covariance or proxy \cite{Zhou2016JSEN,Lee2015JSEN,Zhang2023JSEN}. Learning-based approaches have also been reported for DoA/mmWave sensing, offering robustness under mismatch but requiring training data and providing less transparent performance links to array geometry \cite{Zheng2024JSEN,Su2023JSEN,Rai2021JSEN}. Hybrid beamforming and beamspace MIMO designs optimize precoding and combining matrices to maximize communication metrics such as spectral efficiency or BER under RF-chain constraints \cite{Leyva2024JSEN}; these frameworks select beams to maximize received signal power at each data stream, and the resulting beam patterns can be arbitrary. The present work addresses a structurally different objective: beam selection for receiver-side DoA \emph{estimation} accuracy under a strict DFT beam budget. Specifically, beamspace ESPRIT requires the selected beams to form a \emph{contiguous, shift-invariant} block within each angular sector~\cite{ApertureReduction_Beamspace}; arbitrary power-maximizing beam patterns that violate this contiguity constraint break the shift-invariance equation on which ESPRIT relies. Consequently, the optimization criterion, structural beam constraints, and performance metrics of the proposed framework are fundamentally different from those of beamspace MIMO precoder design.

Despite these advances, there remains a gap between the theoretical capabilities of fully digital beamspace ESPRIT and the practical constraints of HAD mmWave sensor arrays with a limited beam budget. In particular, existing methods for hybrid arrays either (i) assume access to many training configurations or a dense set of beams, (ii) rely on computationally intensive sparse or learning-based estimators, or (iii) treat beam selection heuristically---targeting received power or communication throughput rather than the contiguity and shift-invariance constraints required by beamspace ESPRIT---without exploiting analytical MSE expressions or the Toeplitz structure of the sensor covariance. As a result, there is a need for low-complexity, analytically transparent schemes that (a) respect the contiguity and shift-invariance requirements of beamspace Unitary ESPRIT, (b) operate under strict RF-chain and DFT beam constraints, and (c) bridge much of the performance gap to the Cram\'er--Rao bound for the array geometry and signal model considered here. For hybrid architectures, \cite{ZhangShimWu2022} derives DoA estimation bounds under general analog combining constraints across fully-connected, sub-connected, and switch-based structures, providing a theoretical performance floor for HAD receiver design; the present work complements this analysis by targeting the beam-selection and covariance-denoising problem that arises when the receiver is constrained to a small contiguous DFT beam subset.

In this paper, we develop a covariance-guided DFT beam selection framework for hybrid mmWave sensor arrays tailored to beamspace ESPRIT. A short training phase realizes a virtual centro-symmetric subarray; forward--backward averaging, nonnegative least squares (NNLS) power/noise fitting, and Toeplitz--PSD projection reconstruct a denoised full-aperture covariance. This covariance selects, within each coarse sector, small contiguous DFT-beam blocks under an explicit beam budget, and a sparse beamspace Unitary ESPRIT stage then refines the estimates using only available adjacent beam pairs. In the ideal unitary-DFT setting studied in the main simulations, the fine-stage beam selector reduces to a covariance-guided contiguous energy-window rule. The broader score formulation is retained because it also covers non-unitary effective beam dictionaries arising from phase quantization, codebook oversampling, or mutual coupling.

Our contributions can be summarized as follows.
\begin{itemize}
  \item A covariance reconstruction method for hybrid arrays: virtual centro-symmetric subarray + FBA + NNLS fitting + Toeplitz--PSD projection to obtain a denoised full-aperture covariance.
  \item A sector-wise contiguous DFT-beam selector driven by this covariance, compatible with the shift-invariance/contiguity requirements of beamspace Unitary ESPRIT under strict beam budgets. For the unitary DFT codebook considered in the main experiments, the selector reduces to choosing the maximum-energy contiguous beam block of the denoised beamspace covariance within each sector; the general formulation additionally accommodates non-unitary effective beam dictionaries.
  \item A coarse-to-fine ESPRIT pipeline with an extensive Monte Carlo study (RMSE, failure rate, gap-to-CRB) showing improved accuracy--complexity trade-offs over a sectorization baseline using the same codebook/estimator.
\end{itemize}

The rest of the paper is organized as follows. Section~\ref{sec:system_model} describes the system model, the hybrid training architecture, and the beamspace representation. Section~\ref{sec:coarse_fine_estimation} presents the covariance-guided coarse-to-fine beamspace ESPRIT framework, including the virtual subarray construction, covariance fitting, and DFT beam selection algorithm. Section~\ref{sec:num_results} provides numerical results that compare the proposed method with a sectorization-based beamspace ESPRIT baseline and examine the impact of key design parameters. Section~\ref{sec:conclusion} concludes the paper and outlines directions for future work, including extensions to wideband, two-dimensional, and near-field sensor arrays.

\section{System Model and Hybrid Training Architecture}
\label{sec:system_model}

We summarize the array model, hybrid combining architecture, and DFT beamspace representation. The receiver first acquires low-dimensional hybrid snapshots $\mathbf{y}(n)=\mathbf{W}_{\mathrm{BB}}^{\mathsf{H}} \mathbf{W}_{\mathrm{RF}}^{\mathsf{H}} \mathbf{x}(n)$, constructs a virtual subarray covariance for coarse FBA-TLS-ESPRIT, then uses Toeplitz–PSD covariance fitting to guide the selection of a small contiguous set of DFT beams for fine beamspace ESPRIT. We interpret $\mathbf{x}(n)$ as the sensor-array measurement at probing interval $n$, obtained with a constrained hybrid front-end; the DoA parameters $\boldsymbol{\mu}$ represent target or source angles to be estimated from these constrained measurements.

\subsection{Array and Channel Model}

We consider a uniform linear array (ULA) of
\(M\) antenna elements at the receiver, with inter-element spacing
\(d_{\text{ant}}\) normalized to the carrier wavelength \(\lambda\). The array operates in a narrowband regime and can be interpreted as observing a single OFDM subcarrier during a probing interval. At snapshot index \(n\), the \(M \times 1\) array output vector is modeled as
\begin{equation}
    \mathbf{x}(n) = \mathbf{A}(\boldsymbol{\mu}) \mathbf{s}(n) + \mathbf{w}(n),
    \label{eq:array_output}
\end{equation}
where
\begin{itemize}
  \item \(\mu_k = \tfrac{2\pi d_{\text{ant}}}{\lambda}\sin\theta_k\) is the spatial frequency associated with azimuth angle \(\theta_k\), and \(\boldsymbol{\mu} = [\mu_1, \ldots, \mu_d]^{\mathsf{T}}\) collects the spatial frequencies of the \(d\) far-field paths,
  \item \(\mathbf{s}(n) = [s_1(n), \ldots, s_d(n)]^{\mathsf{T}} \in \mathbb{C}^{d}\) contains the complex path gains at snapshot \(n\),
  \item \(\mathbf{w}(n) \sim \mathcal{CN}(\mathbf{0}, N_0 \mathbf{I}_M)\) denotes spatially white receiver noise, and
  \item \(\mathbf{A}(\boldsymbol{\mu}) = [\mathbf{a}(\mu_1), \ldots, \mathbf{a}(\mu_d)] \in \mathbb{C}^{M \times d}\) is the array steering matrix.
\end{itemize}
For a ULA with element indices \(m = 0,\ldots,M-1\), the steering vector at spatial frequency \(\mu\) is
\begin{equation}
    \mathbf{a}(\mu) =
    \begin{bmatrix}
        1,
        e^{\mathrm{j}\mu},
        \ldots,
        e^{\mathrm{j}(M-1)\mu}
    \end{bmatrix}^{\mathsf{T}}.
    \label{eq:steering_vector}
\end{equation}
We assume throughout that the array is calibrated and that the far-field, narrowband conditions hold over the considered bandwidth and angular sector.

Let \(\mathbf{R}_{\mathbf{s}} = \mathbb{E}[\mathbf{s}(n)\mathbf{s}^{\mathsf{H}}(n)]\) denote the source covariance matrix and \(\mathbf{R}_{\mathbf{x}} = \mathbb{E}[\mathbf{x}(n)\mathbf{x}^{\mathsf{H}}(n)]\) the array output covariance. Under the model \eqref{eq:array_output}, we have
\begin{equation}
    \mathbf{R}_{\mathbf{x}} = \mathbf{A}(\boldsymbol{\mu}) \mathbf{R}_{\mathbf{s}} \mathbf{A}^{\mathsf{H}}(\boldsymbol{\mu}) + N_0 \mathbf{I}_M.
    \label{eq:array_covariance}
\end{equation}
In the numerical results we focus on uncorrelated sources with
\(\mathbf{R}_{\mathbf{s}} = \operatorname{diag}(p_1, \ldots, p_d)\), but the proposed covariance-fitting framework can be extended to more general \(\mathbf{R}_{\mathbf{s}}\).

A convenient scalar performance parameter is the array signal-to-noise ratio (ASNR), defined as
\begin{equation}
    \mathrm{ASNR} \triangleq
    \frac{\mathrm{tr}(\mathbf{A}(\boldsymbol{\mu}) \mathbf{R}_{\mathbf{s}} \mathbf{A}^{\mathsf{H}}(\boldsymbol{\mu}))}{M N_0}
    = \frac{\sum_{\ell=1}^{d} p_\ell \, \|\mathbf{a}(\mu_\ell)\|_2^2}{M N_0}.
    \label{eq:asnr_definition}
\end{equation}
The ASNR captures the average per-antenna SNR and will be used to parameterize the Monte Carlo simulations.

\subsection{Hybrid Combining and DFT Beamspace Representation}
\label{subsec:HBA}
The receiver employs a hybrid analog/digital combining architecture with \(N_{\mathrm{RF}} < M\) RF chains. The analog RF combiner is implemented using a fully connected phase-shifter network,
\begin{equation}
    \mathbf{W}_{\mathrm{RF}} \in \mathbb{C}^{M \times N_{\mathrm{RF}}},
    \label{eq:wrf_def}
\end{equation}
whose entries satisfy a constant-modulus constraint
\(|[\mathbf{W}_{\mathrm{RF}}]_{m,r}| = 1/\sqrt{M}\). We assume continuous-phase shifters for simplicity. In the main numerical study we retain this idealized continuous-phase model so that the impact of covariance reconstruction and beam-budget\-ted selection can be isolated from hardware quantization effects. For a $B$-bit phase shifter with phase resolution $\Delta\varphi = 2\pi/2^B$, the realized analog combining vector departs from the intended DFT column $\mathbf{b}_k$, yielding an effective beam dictionary $\widetilde{\mathbf{B}}$ whose columns are quantization-perturbed DFT vectors. In this regime, the Gram matrix $\mathbf{G}_g(\mathcal{S}_g) = \widetilde{\mathbf{B}}_g^{\mathsf{H}}(\mathcal{S}_g) \widetilde{\mathbf{B}}_g(\mathcal{S}_g) \neq \mathbf{I}$ for generic contiguous subsets, so the conditioning regularization in \eqref{eq:preservationScore}--\eqref{eq:score_def} becomes active and provides an inherent guard against ill-conditioned beam subsets under moderate phase-quantization non-ideality. The severity of this perturbation decreases rapidly with $B$: for $B \ge 4$ the phase error per element is bounded by $\pi/2^B \le 11.25^{\circ}$, and the induced column-wise $\ell_2$ perturbation of $\widetilde{\mathbf{B}}$ relative to $\mathbf{B}$ satisfies $\|\tilde{\mathbf{b}}_k - \mathbf{b}_k\|_2 \le 2\sin(\pi/2^{B+1}) \le 0.20$ per unit-norm column, characterizing the departure from the unitary regime in hardware terms. The end-to-end impact on DoA estimation accuracy is quantified by the Monte Carlo study in Section~\ref{subsec:phase_quant}. The digital baseband combiner,
\begin{equation}
    \mathbf{W}_{\mathrm{BB}} \in \mathbb{C}^{N_{\mathrm{RF}} \times N_{\mathrm{b,slot}}},
    \label{eq:wbb_def}
\end{equation}
is implemented in the digital domain, does not obey modulus constraints, and can be reconfigured across training slots.

At snapshot \(n\), the \(N_{\mathrm{b,slot}} \times 1\) baseband observation is
\begin{equation}
    \mathbf{y}(n) = \mathbf{W}_{\mathrm{BB}}^{\mathsf{H}} \mathbf{W}_{\mathrm{RF}}^{\mathsf{H}} \mathbf{x}(n)
    \in \mathbb{C}^{N_{\mathrm{b,slot}}}.
    \label{eq:hybrid_observation}
\end{equation}
Stacking \(N_{\mathrm{snap}}\) snapshots, we obtain the baseband data matrix
\begin{equation}
    \mathbf{Y} =
    \begin{bmatrix}
        \mathbf{y}(1) & \cdots & \mathbf{y}(N_{\mathrm{snap}})
    \end{bmatrix}
    \in \mathbb{C}^{N_{\mathrm{b,slot}} \times N_{\mathrm{snap}}}.
    \label{eq:baseband_data}
\end{equation}

To exploit the angular sparsity of mmWave channels and to connect with beamspace ESPRIT, we now introduce the DFT beamspace representation. We index the DFT grid over spatial frequency as
\begin{equation}
    \gamma_k \triangleq -\pi + (k-1)\Delta_\mu,
    \qquad
    \Delta_\mu \triangleq \frac{2\pi}{M},
    \qquad
    k=1,\ldots,M.
    \label{eq:dft_grid}
\end{equation}
The unitary (FFT-shifted) DFT beamforming matrix is then defined entry-wise by
\begin{equation}
    [\mathbf{B}]_{m,k} \triangleq \frac{1}{\sqrt{M}} e^{-\mathrm{j}(m-1)\gamma_k},
    \qquad m,k=1,\ldots,M,
    \label{eq:dft_matrix}
\end{equation}
whose columns correspond to fixed beams pointing towards a uniform grid of virtual angles. We use this basis to define a DFT beamspace representation of the array outputs.

In the fully digital case, the beamspace data matrix would be \(\mathbf{B}^{\mathsf{H}} \mathbf{X}\), where \(\mathbf{X} = [\mathbf{x}(1), \ldots, \mathbf{x}(N_{\mathrm{snap}})]\) collects the array outputs. In the hybrid case, the RF combiner \(\mathbf{W}_{\mathrm{RF}}\) is constructed from (possibly phase-shifted) columns of \(\mathbf{B}\), so that each RF chain corresponds to one or a small number of effective DFT beams. The digital combiner \(\mathbf{W}_{\mathrm{BB}}\) then forms suitable linear combinations of these RF-chain outputs.

We denote by \(N_{\mathrm{b}}\) the total number of effective DFT beams that are observed during the training phase and subsequently used in the fine ESPRIT stage. After appropriate re-indexing and scaling, the corresponding beamspace data matrix is
\begin{equation}
    \mathbf{Y}_{\mathrm{b}} \in \mathbb{C}^{N_{\mathrm{b}} \times N_{\mathrm{snap}}},
    \label{eq:beamspace_data}
\end{equation}
and its sample covariance is
\begin{equation}
    \widehat{\mathbf{R}}_{\mathrm{b}} =
    \frac{1}{N_{\mathrm{snap}}}
    \mathbf{Y}_{\mathrm{b}} \mathbf{Y}_{\mathrm{b}}^{\mathsf{H}}.
    \label{eq:rb_sample}
\end{equation}
The central goal of the next section is to use \(\widehat{\mathbf{R}}_{\mathrm{b}}\) to reconstruct a denoised full-aperture covariance and to design a covariance-guided selection of contiguous DFT beams that is suitable for beamspace ESPRIT.

\subsection{Virtual Subarray Training and Practical Realizability}

The coarse FBA-TLS-ESPRIT stage in Section~\ref{sec:coarse_fine_estimation} operates on the covariance of a virtual centro-symmetric contiguous subarray in element space. During a short training phase, we reconfigure the hybrid combiner across a small number of slots and linearly post-process the resulting baseband measurements to emulate access to this subarray. In the analysis and simulations we adopt an idealized setting in which this effective mapping is accurate enough that the coarse-stage covariance can be formed as if the subarray samples were observed directly. The concrete construction of the effective mapping and the associated digital combiner design are described in Section~\ref{subsec:coarse_estimation}, and the subsequent covariance regularization is described in Section~\ref{subsec:covariance_fitting}.

\paragraph*{Training overhead and feasibility}
Let $N_{\mathrm{b,slot}}$ denote the number of digital outputs per training slot in \eqref{eq:hybrid_observation}. If $N_{\mathrm{b,slot}}=N_{\mathrm{RF}}$, we use a single hybrid configuration in the coarse stage: we fix $\mathbf{W}_{\mathrm{RF}}$ and compute an effective de-mixing matrix $\widetilde{\mathbf{W}}_{\mathrm{BB}}\in\mathbb{C}^{N_{\mathrm{RF}}\times N_{\mathrm{RF}}}$ once from \eqref{eq:WBB_LS}, then collect $N_{\mathrm{snap}}$ snapshots with the same combiner. If $N_{\mathrm{b,slot}}<N_{\mathrm{RF}}$, we use $L=\lceil N_{\mathrm{RF}}/N_{\mathrm{b,slot}}\rceil$ slots with different $\mathbf{W}_{\mathrm{BB}}^{(\ell)}$ and stack the outputs to obtain an effective $N_{\mathrm{RF}}$-dimensional observation for forming the coarse covariance. The mapping is feasible whenever $\mathbf{J}_{\mathcal{M}}\mathbf{W}_{\mathrm{RF}}$ has full column rank. When this matrix is ill-conditioned, the ridge term $\alpha_{\mathrm{LS}}$ in \eqref{eq:WBB_LS} limits noise amplification, and the remaining mismatch can be absorbed into the coarse-stage noise model.

\section{Covariance-Guided Coarse-to-Fine Beamspace ESPRIT}
\label{sec:coarse_fine_estimation}
\subsection{Coarse ESPRIT on a Virtual Element-Space Subarray}
\label{subsec:coarse_estimation}

We begin by defining the indices of the virtual subarray. We select a centro-symmetric block of $N_{\mathrm{RF}}$ antennas around the array center:
\begin{equation}
    \mathcal{M}
    \triangleq
    \left\{ m \in \{1,\ldots,M\} :
    \left| m - \frac{M+1}{2} \right|
    \le \frac{N_{\mathrm{RF}}-1}{2} \right\}.
    \label{eq:centro-symmetricMask}
\end{equation}
We use 0-based indexing in the steering-vector phase progression and 1-based indexing for set notation; they are consistent up to a shift. Let $\mathbf{J}_{\mathcal{M}}\in\{0,1\}^{N_{\mathrm{RF}}\times M}$ denote the corresponding selection matrix, so that the virtual subarray data matrix is
\begin{equation}
  \mathbf{X}_{\mathcal{M}}
  \triangleq
  \mathbf{J}_{\mathcal{M}}\mathbf{X}
  \in \mathbb{C}^{N_{\mathrm{RF}}\times N_{\mathrm{snap}}}.
  \label{eq:virtual_subarray_data}
\end{equation}

In a hybrid receiver the subarray cannot be observed directly. For a given $\mathbf{W}_{\mathrm{RF}}$, we define an
\emph{effective} digital de-mixing matrix $\widetilde{\mathbf{W}}_{\mathrm{BB}}\in\mathbb{C}^{N_{\mathrm{RF}}\times N_{\mathrm{RF}}}$ such that $\mathbf{J}_{\mathcal{M}}\mathbf{W}_{\mathrm{RF}}\widetilde{\mathbf{W}}_{\mathrm{BB}}\approx \mathbf{I}_{N_{\mathrm{RF}}}$, i.e., after the required slot-stacking (post-processing) the resulting matrix $\mathbf{Y}_{\mathrm{coarse}}$ approximates $\mathbf{X}_{\mathcal{M}}$. If $N_{\mathrm{b,slot}}=N_{\mathrm{RF}}$, we set $\mathbf{Y}_{\mathrm{coarse}} \triangleq \mathbf{Y}$. If $N_{\mathrm{b,slot}}<N_{\mathrm{RF}}$, we stack the $L$ slot outputs into $\widetilde{\mathbf{Y}}\in\mathbb{C}^{N_{\mathrm{RF}}\times N_{\mathrm{snap}}}$ and, for notational simplicity, we denote this stacked matrix by $\mathbf{Y}_{\mathrm{coarse}}$.

When $\mathbf{J}_{\mathcal{M}}\mathbf{W}_{\mathrm{RF}}$ is well conditioned we set
$\widetilde{\mathbf{W}}_{\mathrm{BB}}=(\mathbf{J}_{\mathcal{M}}\mathbf{W}_{\mathrm{RF}})^{-1}$; otherwise we use the regularized LS solution:
\begin{equation}
    \widetilde{\mathbf{W}}_{\mathrm{BB}}
    = \arg\min_{\mathbf{Z}\in\mathbb{C}^{N_{\mathrm{RF}}\times N_{\mathrm{RF}}}}
    \big\| \mathbf{J}_{\mathcal{M}}\mathbf{W}_{\mathrm{RF}}\mathbf{Z} - \mathbf{I}_{N_{\mathrm{RF}}} \big\|_{F}^{2}
    + \alpha_{\mathrm{LS}} \|\mathbf{Z}\|_{F}^{2},
    \label{eq:WBB_LS}
\end{equation}
with a small regularization parameter $\alpha_{\mathrm{LS}}>0$, computed once offline.

Given the approximate relation, we form the sample covariance of the virtual subarray as
\begin{equation}
     \widehat{\mathbf{R}}_{\mathcal{M}}
     \triangleq
     \frac{1}{N_{\mathrm{snap}}}\mathbf{X}_{\mathcal{M}}\mathbf{X}_{\mathcal{M}}^{\mathsf{H}}
     \approx
     \frac{1}{N_{\mathrm{snap}}}\mathbf{Y}_{\mathrm{coarse}}\mathbf{Y}_{\mathrm{coarse}}^{\mathsf{H}}
     \label{eq:R_M_sample}
\end{equation}
and we apply forward--backward averaging to exploit the centro-symmetry of $\mathcal{M}$ and improve robustness to correlated paths
\begin{equation}
    \widehat{\mathbf{R}}_{\mathrm{FBA}}
    =
    \frac{1}{2}\big[
      \widehat{\mathbf{R}}_{\mathcal{M}}
      +
      \mathbf{\Pi}_{N_{\mathrm{RF}}}\widehat{\mathbf{R}}_{\mathcal{M}}^{*}\mathbf{\Pi}_{N_{\mathrm{RF}}}
    \big],
    \label{eq:fbAveCov}
\end{equation}
where $\mathbf{\Pi}_{N_{\mathrm{RF}}}$ is the $N_{\mathrm{RF}}\times N_{\mathrm{RF}}$ exchange (reversal) matrix. Let
\[
    \widehat{\mathbf{R}}_{\mathrm{FBA}}
    =
    \widehat{\mathbf{U}}\widehat{\mathbf{\Lambda}}\widehat{\mathbf{U}}^{\mathsf{H}}
\]
denote the eigendecomposition of $\widehat{\mathbf{R}}_{\mathrm{FBA}}$, and let $\widehat{\mathbf{U}}_{\mathrm{S}}\in\mathbb{C}^{N_{\mathrm{RF}}\times d}$ contain the $d$ principal eigenvectors associated with the largest eigenvalues.

We apply TLS-ESPRIT~\cite{TensorESPRIT_Beamspace,Asilomar_ESPRIT_MSE} to $\widehat{\mathbf{R}}_{\mathrm{FBA}}$: the signal subspace $\widehat{\mathbf{U}}_{\mathrm{S}}$ is projected onto two overlapping shifted subarrays, and the coarse spatial-frequency estimates $\widehat{\boldsymbol{\mu}}_{\mathrm{coarse}} \in \mathbb{R}^d$ are extracted from the eigenvalues of the resulting rotation matrix $\widehat{\boldsymbol{\Psi}}$.

\subsection{Covariance Fitting and Toeplitz--PSD Projection}
\label{subsec:covariance_fitting}

The coarse FBA-TLS-ESPRIT stage described above provides spatial-frequency estimates $\widehat{\boldsymbol{\mu}}_{\mathrm{coarse}} \in \mathbb{R}^{d}$ based on the virtual subarray covariance
$\widehat{\mathbf{R}}_{\mathcal{M}}$ in \eqref{eq:R_M_sample}. In this subsection we use $\widehat{\boldsymbol{\mu}}_{\mathrm{coarse}}$ to fit a structured covariance model and to reconstruct a denoised full-aperture covariance matrix, which will later guide the DFT beam selection and the fine beamspace ESPRIT stage.

We adopt the standard uncorrelated-source model
\begin{equation}
  \mathbf{R}_{\mathbf{x}}
  \triangleq
  \mathbb{E}[\mathbf{x}(n)\mathbf{x}^{\mathsf{H}}(n)]
  =
  \mathbf{A}(\boldsymbol{\mu}) \mathbf{R}_{\mathbf{s}} \mathbf{A}^{\mathsf{H}}(\boldsymbol{\mu})
  + N_0\mathbf{I}_{M},
  \label{eq:cov_model_uncorr}
\end{equation}
with
$\mathbf{R}_{\mathbf{s}} = \operatorname{diag}(p_{1},\ldots,p_{d}) \succeq \mathbf{0}$ and noise variance $N_0 \ge 0$.
For a ULA, $\mathbf{R}_{\mathbf{x}}$ is Hermitian Toeplitz whenever the propagation environment is
wide-sense stationary across the aperture. Under this model, the virtual subarray covariance is
\begin{equation}
   \mathbf{R}_{\mathcal{M}}
   =
   \mathbf{J}_{\mathcal{M}} \mathbf{R}_{\mathbf{x}} \mathbf{J}_{\mathcal{M}}^{\mathsf{H}}
   =
   \mathbf{A}_{\mathcal{M}}(\boldsymbol{\mu}) \mathbf{R}_{\mathbf{s}}
   \mathbf{A}_{\mathcal{M}}^{\mathsf{H}}(\boldsymbol{\mu})
   + N_0\mathbf{I}_{N_{\mathrm{RF}}},
   \label{eq:cov_model_subarray}
\end{equation}
where
$\mathbf{A}_{\mathcal{M}}(\boldsymbol{\mu})
 \triangleq \mathbf{J}_{\mathcal{M}}\mathbf{A}(\boldsymbol{\mu}) \in \mathbb{C}^{N_{\mathrm{RF}}\times d}$.
In practice, we do not know $\boldsymbol{\mu}$ or $\mathbf{R}_{\mathbf{s}}$, and we only observe the
sample covariance of the hybrid outputs, which we approximate as
$\widehat{\mathbf{R}}_{\mathcal{M}}$ in \eqref{eq:R_M_sample} after forward--backward averaging.

Given the coarse FBA-TLS-ESPRIT estimates
$\widehat{\boldsymbol{\mu}}_{\mathrm{coarse}}$, we construct the steering matrices
\begin{IEEEeqnarray}{rCl}
  \mathbf{A}_{\mathcal{M}}
  &\triangleq&
  \mathbf{A}_{\mathcal{M}}(\widehat{\boldsymbol{\mu}}_{\mathrm{coarse}})
  \in \mathbb{C}^{N_{\mathrm{RF}} \times d}, \nonumber \\
  \mathbf{A}_{\mathrm{full}}
  &\triangleq&
  \mathbf{A}(\widehat{\boldsymbol{\mu}}_{\mathrm{coarse}})
  \in \mathbb{C}^{M \times d}.
  \label{eq:A_M_A_full}
\end{IEEEeqnarray}
We then estimate the path powers and noise variance by solving a small nonnegative
least-squares problem that fits the virtual subarray covariance model
\eqref{eq:cov_model_subarray} to the sample covariance $\widehat{\mathbf{R}}_{\mathcal{M}}$.
Let
\begin{IEEEeqnarray}{rCl}
    \mathbf{v}_{\mathcal{M}}
    &\triangleq& \mathrm{vec}\big(\widehat{\mathbf{R}}_{\mathcal{M}}\big),
    \nonumber \\
    \mathbf{G}_{\mathcal{M}}
    &\triangleq&
    \begin{bmatrix}
      \mathrm{vec}\!\big(\mathbf{a}_{\mathcal{M},1}\mathbf{a}_{\mathcal{M},1}^{\mathsf{H}}\big) &
      \cdots &
      \mathrm{vec}\!\big(\mathbf{a}_{\mathcal{M},d}\mathbf{a}_{\mathcal{M},d}^{\mathsf{H}}\big) &
      \mathrm{vec}(\mathbf{I}_{N_{\mathrm{RF}}})
    \end{bmatrix},
    \label{eq:G_M_def}
\end{IEEEeqnarray}
where $\mathbf{a}_{\mathcal{M},\ell}$ denotes the $\ell$th column of
$\mathbf{A}_{\mathcal{M}}$.
Under the uncorrelated-source model,
$\mathbf{v}_{\mathcal{M}}$ is approximately equal to $\mathbf{G}_{\mathcal{M}}\boldsymbol{\xi}$ with
\begin{equation}
    \boldsymbol{\xi}
    \triangleq
    \begin{bmatrix}
        p_{1} & \cdots & p_{d} & N_0
    \end{bmatrix}^{\mathsf{T}}.
\end{equation}
We estimate $\boldsymbol{\xi}$ via
\begin{equation}
    \widehat{\boldsymbol{\xi}}
    =
    \arg\min_{\mathbf{z}\in\mathbb{R}_+^{d+1}}
    \big\| \mathbf{G}_{\mathcal{M}}\mathbf{z} - \mathbf{v}_{\mathcal{M}} \big\|_{2}^{2},
    \label{eq:nnls_covfit}
\end{equation}
which is a $(d+1)$-dimensional nonnegative least-squares problem that can be solved
efficiently. The first $d$ components of $\widehat{\boldsymbol{\xi}}$ yield the power estimates
$\widehat{p}_{1},\ldots,\widehat{p}_{d}$, and the last component yields
$\widehat{N}_0$. In the simulations we use a standard NNLS solver with a small
regularization term to improve numerical conditioning.

From the estimated powers, we reconstruct a full-aperture signal covariance as
\begin{equation}
    \widehat{\mathbf{R}}_{\mathrm{sig}}
    \triangleq
    \mathbf{A}_{\mathrm{full}}
    \, \widehat{\mathbf{P}} \,
    \mathbf{A}_{\mathrm{full}}^{\mathsf{H}},
    \qquad
    \widehat{\mathbf{P}}
    \triangleq
    \operatorname{diag}(\widehat{p}_{1},\ldots,\widehat{p}_{d}),
    \label{eq:signal_cov_recon}
\end{equation}
and we form the preliminary full-aperture covariance
\begin{equation}
    \widehat{\mathbf{R}}_{\mathrm{full}}
    \triangleq
    \widehat{\mathbf{R}}_{\mathrm{sig}} + \widehat{N}_0\mathbf{I}_{M}.
    \label{eq:full_cov_pre}
\end{equation}
This covariance is consistent with the uncorrelated-source model
and the coarse ESPRIT estimates, but it is affected by sampling noise,
coarse-estimation errors, and any mismatch between the virtual subarray
model \eqref{eq:cov_model_subarray} and the true hybrid measurements.

To further denoise $\widehat{\mathbf{R}}_{\mathrm{full}}$ and to enforce the physical
structure of a calibrated ULA, we project it onto the cone of Hermitian Toeplitz
positive-semidefinite (PSD) matrices. Specifically, we solve
\begin{IEEEeqnarray}{rCl}
   \widehat{\mathbf{R}}_{\mathrm{T}}
   = &&
   \arg\min_{\mathbf{T}}
   \big\| \mathbf{T} - \widehat{\mathbf{R}}_{\mathrm{full}} \big\|_{F}^{2} \nonumber \\
   &&\text{subject to} \quad
   \mathbf{T} = \mathbf{T}^{\mathsf{H}},\
   \mathbf{T}\ \text{Toeplitz},\
   \mathbf{T} \succeq \mathbf{0},
   \label{eq:toeplitz_psd_projection}
\end{IEEEeqnarray}
where $\|\cdot\|_{F}$ denotes the Frobenius norm. The Toeplitz constraint encodes spatial wide-sense stationarity across the ULA, and the PSD constraint guarantees that the result is a valid covariance matrix. In practice, the optimization \eqref{eq:toeplitz_psd_projection} reduces to a small real-valued quadratic program in the first column of $\mathbf{T}$. The real-valued quadratic-program reformulations and implementation details for the NNLS fit and the Toeplitz--PSD projection are provided in Appendix~\ref{app:qp_real}. We refer to $\widehat{\mathbf{R}}_{\mathrm{T}}$ as the \emph{Toeplitz--PSD projected covariance}.

This projection reduces estimation variance by enforcing Toeplitz structure and guarantees a valid covariance via PSD, improving robustness at finite $N_{\mathrm{snap}}$ and under moderate mismatch.

In the remainder of the paper, we use $\widehat{\mathbf{R}}_{\mathrm{T}}$ as the
full-aperture covariance underlying the DFT beam scoring and selection in
Section~\ref{sec:coarse_fine_estimation}, as well as in the performance
interpretation based on beamspace ESPRIT MSE expressions.

\subsection{Covariance-Guided DFT Beam Selection}
\label{subsec:cov_beam_selection}

The Toeplitz--PSD projected covariance obtained in
Section~\ref{subsec:covariance_fitting} provides detailed information about
how the signal energy is distributed over array lags and, after transformation
to DFT beamspace, over virtual angles. In this subsection we use this
covariance to select, under a strict beam budget, small contiguous subsets of
DFT beams that (i) concentrate signal energy around the dominant paths,
(ii) preserve a large effective aperture, and (iii) satisfy the contiguity
constraints required by beamspace Unitary ESPRIT. The resulting beam subsets
define the fine-stage hybrid combiner and thus the mmWave sensor front-end
used for high-resolution DoA estimation.

Let $\widehat{\mathbf{R}}_{\mathrm{T}}$ denote the Toeplitz--PSD projected
full-aperture covariance from \eqref{eq:toeplitz_psd_projection}, and let
$\widehat{N}_0$ be the noise variance estimate from \eqref{eq:nnls_covfit}.
We define the denoised signal covariance as
\begin{equation}
    \widetilde{\mathbf{R}}_{\mathrm{sig}}
    \triangleq
    \widehat{\mathbf{R}}_{\mathrm{T}} - \widehat{N}_0 \mathbf{I}_M,
    \label{eq:Rs_tilde_def}
\end{equation}
which enforces Hermitian Toeplitz structure and removes the
estimated spatially white noise component. This matrix summarizes both the
coarse spatial-frequency estimates and the fitted path powers, after
structural denoising.

\subsubsection{Sectorization and sector-wise beam pools}

We adopt a sliding-window sectorization of the DFT grid similar
to~\cite{TensorESPRIT_Beamspace}. The phase-shifted DFT beamforming matrix
\[
    \mathbf{B}
    =
    [\mathbf{b}_1,\ldots,\mathbf{b}_M]
    \in\mathbb{C}^{M\times M},
\]
introduced in Section~\ref{subsec:HBA}, defines $M$ fixed beams pointing
towards a uniform grid of virtual angles. Let $\gamma_k$ (with spacing $\Delta_\mu$) be defined as in \eqref{eq:dft_grid} and denote the spatial frequency of the $k$th DFT beam. Using the coarse FBA-TLS-ESPRIT estimates $\widehat{\boldsymbol{\mu}}_{\mathrm{coarse}}$ and a user-specified number of sectors $G$, we construct $G$ overlapping angular sectors that cover the signal-of-interest (SoI) region. For sector $g$ we obtain a sorted index set
\[
    \mathcal{B}_g \subset \{1,\ldots,M\}, \quad g=1,\ldots,G,
\]
where $\mathcal{B}_g$ contains the DFT beam indices whose main lobes lie in
the $g$-th sector. The sets $\{\mathcal{B}_g\}_{g=1}^{G}$ form sector-wise
beam pools from which we will select the fine-stage beams. This construction
ensures that, within each sector, the candidate beams are roughly aligned with
the local support of the dominant paths.

A coarse-angle error affects the fine stage only through the sector definition. Let each sector be formed as a width-$W$ beam window around the corresponding coarse estimate. Then the true path remains inside the sector beam pool whenever $|\widehat{\mu}_{\ell,\mathrm{coarse}}-\mu_\ell|\le (W/2)\Delta_{\mu}$. Under this condition the fine-stage search does not exclude the beams that maximize the response of the $\ell$-th path; larger errors may shift the window and increase the failure probability.

\subsubsection{Candidate beam sets and covariance-capture score}

Within each sector $\mathcal{B}_g$, we seek one contiguous block of
$K_g \ge 2$ DFT beams
\[
    \mathcal{S}_g
    =
    \{\kappa_j,\kappa_{j+1},\ldots,\kappa_{j+K_g-1}\}
    \subset \mathcal{B}_g,
\]
that is most informative for the sources in the $g$-th sector.
For a given candidate set $\mathcal{S}_g$, we collect the corresponding
beamformers into
\[
    \mathbf{B}_g(\mathcal{S}_g)
    \triangleq
    \big[
        \mathbf{b}_{\kappa_j},
        \mathbf{b}_{\kappa_{j+1}},
        \ldots,
        \mathbf{b}_{\kappa_{j+K_g-1}}
    \big]
    \in \mathbb{C}^{M\times K_g},
\]
and define the associated Gram matrix
\[
    \mathbf{G}_g(\mathcal{S}_g)
    \triangleq
    \mathbf{B}_{g}^{\mathsf{H}}(\mathcal{S}_g)\mathbf{B}_g(\mathcal{S}_g)
    \in \mathbb{C}^{K_g\times K_g}.
\]

The denoised covariance $\widetilde{\mathbf{R}}_{\mathrm{sig}}$ provides a measure of how much signal energy is captured by the subspace spanned by the beams in $\mathcal{S}_g$. Motivated by first-order beamspace ESPRIT MSE expressions that link the error to the effective SNR and beamspace conditioning~\cite{Asilomar_ESPRIT_MSE}, we quantify this via the covariance-capture score
\begin{equation}
    \mathrm{cap}(\mathcal{S}_g)
    =
    \operatorname{tr}\!\Big[
        \big(\mathbf{G}_g(\mathcal{S}_g) + \eta \mathbf{I}_{K_g}\big)^{-1}
        \mathbf{B}_{g}^{\mathsf{H}}(\mathcal{S}_g)
        \widetilde{\mathbf{R}}_{\mathrm{sig}}
        \mathbf{B}_g(\mathcal{S}_g)
    \Big],
    \label{eq:preservationScore}
\end{equation}
where $\eta\geq 0$ is a small Tikhonov parameter that improves numerical
stability and can be interpreted as a regularized least-squares projection
onto the span of $\mathbf{B}_g(\mathcal{S}_g)$.

To discourage highly ill-conditioned beam sets, we also measure the squared
condition number
\[
    \kappa_g^2(\mathcal{S}_g)
    \triangleq
    \operatorname{cond}\big(\mathbf{G}_g(\mathcal{S}_g)\big)^2,
\]
and define the final score as
\begin{equation}
    \mathrm{score}(\mathcal{S}_g)
    =
    \frac{\mathrm{cap}(\mathcal{S}_g)}{1 + \alpha\,\kappa_g^2(\mathcal{S}_g)},
    \label{eq:score_def}
\end{equation}
where $\alpha\geq 0$ trades off signal-energy capture against poor
conditioning. In Section~\ref{sec:num_results} we use $\alpha=0.5$ and $\eta=0$. When $\mathbf{B}$ is unitary, both parameters only scale the score and do not change $\mathcal{S}_g^\star$; see Remark~\ref{rem:unitary_dft}.

\begin{remark}[Unitary DFT]\label{rem:unitary_dft}
    If $\mathbf{B}$ is unitary, then $\mathbf{G}_g(\mathcal{S}_g)=\mathbf{I}$ and \eqref{eq:preservationScore} reduces (up to constant factors) to $\sum_{m\in\mathcal{S}_g}\rho_m$, where $\rho_m$ is defined in \eqref{eq:beamspace_power_profile}.
    \begin{equation}
        \rho_m \triangleq \mathbf{b}_m^{\mathsf{H}}\widetilde{\mathbf{R}}_{\mathrm{sig}}\mathbf{b}_m,\qquad m=1,\ldots,M.
        \label{eq:beamspace_power_profile}
    \end{equation}
    
    Thus, within each sector, selecting $\mathcal{S}_g$ is equivalent to sliding a width-$K_g$ energy window over $\{\rho_m\}$ and choosing the maximum-energy window.
    
    In this unitary case, the Tikhonov term $\eta$ in \eqref{eq:preservationScore} and the conditioning penalty in \eqref{eq:score_def} only scale the score by constants and therefore do not affect the maximizing window $\mathcal{S}_g^\star$. As a result, the conditioning term is inactive in the numerical study of Section~\ref{sec:num_results}, where the beam dictionary is the $M$-point unitary DFT. The performance gain of the proposed selector over sectorization is instead driven by computing the energy profile $\{\rho_m\}$ from the denoised Toeplitz--PSD covariance $\widetilde{\mathbf{R}}_{\mathrm{sig}}$, which suppresses noise and leakage before the window search. In the unitary-DFT case used in the main experiments, the selector is therefore best understood as a denoised contiguous energy-window rule constrained by ESPRIT shift-invariance compatibility; its advantage over pure sectorization lies entirely in the quality of $\widetilde{\mathbf{R}}_{\mathrm{sig}}$, not in the conditioning terms of \eqref{eq:preservationScore}--\eqref{eq:score_def}.
\end{remark}

\subsubsection{Implementation and candidate pruning}

For moderate sector widths, we enumerate all possible contiguous
$K_g$-beam windows inside $\mathcal{B}_g$ and evaluate
$\mathrm{score}(\mathcal{S}_g)$ for each candidate. For wide sectors, this
brute-force enumeration may be unnecessary. In that case we use the
beamspace power profile \eqref{eq:beamspace_power_profile} to prune
candidates before evaluating \eqref{eq:preservationScore}--\eqref{eq:score_def}.
Specifically, we restrict the starting indices $\kappa_j$ to those for which
the average beamspace power over the window,
\[
   \frac{1}{K_g} \sum_{m\in\mathcal{S}_g} \rho_m,
\]
exceeds a small fraction of the maximum $\rho_m$ observed in the sector.
This focuses the search on windows that already carry a significant amount of
signal energy, in line with the energy-window interpretation above.

In each sector we select
\[
    \mathcal{S}_g^\star
    =
    \arg\max_{\mathcal{S}_g\subset\mathcal{B}_g}
    \mathrm{score}(\mathcal{S}_g),
\]
with ties broken in favour of smaller condition number and more central
windows within $\mathcal{B}_g$. The union
\begin{equation}
    \mathcal{K}_{\text{fine}}
    \triangleq
    \bigcup_{g=1}^{G} \mathcal{S}_g^\star
    \subset \{1,\ldots,M\}
    \label{eq:K_fine_def}
\end{equation}
defines the DFT beams used by the fine-stage RF combiner in Section~\ref{subsec:fine_beamspace_esprit}. Pseudocode for the covariance-guided beam selection is provided in Appendix~\ref{app:beam_selection}.

\subsection{Fine Spatial Frequency Estimation via Sparse Beamspace Unitary ESPRIT}
\label{subsec:fine_beamspace_esprit}

In the final stage, we perform high-resolution spatial-frequency estimation
using a sparse beamspace Unitary ESPRIT algorithm that operates only on the
contiguous DFT beams selected in Section~\ref{subsec:cov_beam_selection}. The
key idea is to exploit the shift-invariance structure between adjacent DFT
beams inside each selected block, while keeping the effective beamspace
dimension small so that the overall complexity is dominated by a single
low-dimensional ESPRIT call.

Given the selected beam indices $\mathcal{K}_{\text{fine}}$ in \eqref{eq:K_fine_def}, we construct a fine-stage hybrid combiner whose RF part uses the DFT beams $\{\mathbf{b}_k : k \in \mathcal{K}_{\text{fine}}\}$ and whose digital part routes the corresponding RF-chain outputs to separate baseband channels. The resulting fine-stage beamspace observation is
\begin{equation}
    \mathbf{Y}_{\mathrm{b,fine}}
    \in \mathbb{C}^{N_{\mathrm{RF}}^{\mathrm{fine}} \times N_{\mathrm{snap}}},
    \qquad
    N_{\mathrm{RF}}^{\mathrm{fine}}
    \triangleq |\mathcal{K}_{\text{fine}}|,
    \label{eq:Yb_fine}
\end{equation}
where each row of $\mathbf{Y}_{\mathrm{b,fine}}$ corresponds to one selected DFT beam in $\mathcal{K}_{\text{fine}}$ and $N_{\mathrm{RF}}^{\mathrm{fine}}$ is equal to the total number of selected beams. In contrast to classical beamspace ESPRIT that assumes a consecutive block of DFT beams, here the indices in $\mathcal{K}_{\text{fine}}$ form a union of contiguous blocks (e.g., one block per sector) and may not be globally consecutive.

We apply the standard real-valued Unitary ESPRIT framework~\cite{TensorESPRIT_Beamspace,Asilomar_ESPRIT_MSE,ZhangShimWu2022} to $\mathbf{Y}_{\mathrm{b,fine}}$: the beamspace data is pre-rotated to form a real-valued augmented matrix $\mathbf{Y}_{\mathrm{UE}}$, whose $d$-dimensional signal subspace $\mathbf{U}_{\mathrm{S}}$ is extracted via SVD. Two sparse selection matrices $\mathbf{J}_1^{\mathrm{(b)}},\mathbf{J}_2^{\mathrm{(b)}} \in \{0,1\}^{|\mathcal{P}|\times N_{\mathrm{RF}}^{\mathrm{fine}}}$ are assembled from the set of admissible adjacent beam pairs
\begin{equation}
    \mathcal{P}
    \triangleq
    \big\{(\ell_r,\ell_{r+1}) :
          \ell_{r+1} = \ell_r + 1,\;
          r=1,\ldots,N_{\mathrm{RF}}^{\mathrm{fine}}{-}1\big\},
    \label{eq:pair_set}
\end{equation}
enforcing shift-invariance only across actually contiguous beam indices and discarding cross-sector gaps. The rotation matrix $\widehat{\boldsymbol{\Psi}}$ is estimated from the sparse shift-invariance equation $\mathbf{J}_2^{\mathrm{(b)}}\mathbf{U}_{\mathrm{S}} \approx \mathbf{J}_1^{\mathrm{(b)}}\mathbf{U}_{\mathrm{S}}\boldsymbol{\Psi}$ via least squares (or TLS when $|\mathcal{P}|$ is small), and the fine spatial-frequency estimates $\widehat{\mu}_{k,\mathrm{fine}}$ are recovered from its eigenvalues.

By construction, $\mathbf{J}_1^{\mathrm{(b)}}$ and $\mathbf{J}_2^{\mathrm{(b)}}$ use only those
adjacent DFT beam pairs actually present in $\mathcal{K}_{\text{fine}}$, avoiding extrapolation
across missing beams and ensuring compatibility with strict beam budgets and arbitrary
sector configurations. The complexity is dominated by a single $d$-dimensional
eigendecomposition. In the numerical results we match the paths across Monte Carlo runs
by angular proximity and report RMSE and failure rate for the combined coarse-to-fine
estimates $\widehat{\boldsymbol{\mu}}_{\mathrm{fine}}$.

\section{Numerical Results}
\label{sec:num_results}

\subsection{Experimental Setup}
\label{subsec:exp_setup}
We consider a mmWave sensing scenario with a ULA of $M{=}32$ antennas and inter-element spacing $d_{\mathrm{ant}} = \lambda/2$ and $d{=}3$ uncorrelated propagation paths, which can be interpreted as three dominant paths or targets. The broadside-normalized spatial frequencies are fixed to
\[
    \boldsymbol{\mu}_{\text{true}}
    =
    [-2.1,\,0.5,\,2.5]^{\mathsf{T}}\,\mathrm{rad}
\]
and the corresponding path power vector is
\[
    \mathbf{p}
    =
    [0.95,\,0.5,\,0.1]^{\mathsf{T}}.
\]
This configuration reflects a strongly dominant line-of-sight or main cluster and two weaker paths, as often observed in mmWave massive MIMO channels. We adopt the signal model in Section~\ref{sec:system_model} and vary the array signal-to-noise ratio (ASNR), number of snapshots, and beam budget to assess the performance of the proposed covariance-guided beamspace ESPRIT framework. The complete simulation code is publicly available~\cite{senyuva2026covguided_code}.

Unless stated otherwise, parameters follow Table~\ref{tab:sim_params} ($M=32$, $N_{\mathrm{RF}}=12$, $N_{\mathrm{snap}}=100$, ASNR $-10$--$20$ dB, $10^4$ trials).

\begin{table}[t]
  \centering
  \caption{Simulation and algorithm parameters (unless stated otherwise).}
  \label{tab:sim_params}
  \setlength{\tabcolsep}{2pt}
  \scriptsize
  \begin{tabular}{ll}
    \toprule
    Parameter & Value / Description \\
    \midrule
    Array size &
    $M = 32$ ULA, $d_{\text{ant}} = \lambda/2$ \\

    RF chains &
    $N_{\mathrm{RF}} = 12$ \\

    Paths &
    $d = 3$ uncorrelated Gaussian paths \\

    Snapshots &
    $N_{\mathrm{snap}} = 100$ per trial \\

    ASNR &
    $[-10, 20]$ dB, step $1$ dB \\

    Monte Carlo &
    $10^{4}$ trials per ASNR point \\

    Failure criterion &
    Failure rate defined in \eqref{eq:failureRatedef} \\

    Virtual subarray &
    Centered contiguous block $\mathcal{M}$,
    $|\mathcal{M}| = N_{\mathrm{RF}}$ \\

    Hybrid mapping &
    $\widetilde{\mathbf{W}}_{\mathrm{BB}}$ from LS:
    $\mathbf{J}_{\mathcal{M}}\mathbf{W}_{\mathrm{RF}}\widetilde{\mathbf{W}}_{\mathrm{BB}}
    \approx \mathbf{I}_{N_{\mathrm{RF}}}$ \\

    Sector window &
    $W = 4$ DFT beams \\

    Beams per sector &
    $K_g = 2$ (unless stated otherwise) \\

    Covariance model &
    Toeplitz--PSD projected
    $\widehat{\mathbf{R}}_{\mathrm{T}}$ \\

    Score parameters &
    $\alpha = 0.5$, $\eta=0$ (unitary DFT; selection-invariant) \\

    Beam pruning &
    $\texttt{prune\_topq} = 0$  \\

    RF allocation &
    $N_{\mathrm{RF}}^{\mathrm{coarse}}{=}12 \to N_{\mathrm{RF}}^{\mathrm{fine}}{=}6$ \\
    \bottomrule
  \end{tabular}
\end{table}

We compare the proposed covariance-guided framework against a sectorization-based baseline~\cite{TensorESPRIT_Beamspace} using the same DFT codebook, estimator, and evaluation metrics. Five pipelines are evaluated: (1)~proposed coarse stage (centro-symmetric FBA-TLS-ESPRIT, Section~\ref{subsec:coarse_estimation}); (2)~coarse stage from~\cite{TensorESPRIT_Beamspace} (non-symmetric mask); (3)~proposed fine stage --- (1) followed by covariance-guided $K_g{=}2$ beams per sector and sparse Unitary ESPRIT (Section~\ref{subsec:fine_beamspace_esprit}); (4)~fine stage from~\cite{TensorESPRIT_Beamspace} --- (2) followed by sectorization-based $K_g{=}2$ beams and the same estimator; (5)~oracle beams (true-direction DFT beams as in~\cite{Asilomar_ESPRIT_MSE}).

This $12\to 6$ split is chosen for the following reasons.
The coarse stage requires enough RF chains for the virtual centro-symmetric
subarray to support stable FBA-TLS-ESPRIT with $d{=}3$ paths;
$N_{\mathrm{RF}}{=}12$ provides an effective aperture ratio of
$12/32 = 37.5\%$, which is sufficient for reliable sector assignment at
moderate ASNR while keeping the coarse-stage hardware cost well below that of
a fully digital front end.
The fine stage reduces to $K_g{=}2$ contiguous beams per sector across
$G{=}3$ sectors, giving $N_{\mathrm{RF}}^{\mathrm{fine}}{=}6$: the minimum budget
that supports a valid shift-invariance equation for $d{=}3$ paths under the
Unitary ESPRIT constraint $|\mathcal{P}|\ge d$.
We do not claim that the $12\to 6$ split is globally optimal; the ablation
study in Section~\ref{subsec:ablations} and Fig.~\ref{fig:A1_pareto} of Appendix~\ref{app:ablations} confirm it
lies near the knee of the accuracy--runtime trade-off for the considered
scenarios.

Three evaluation metrics are used. The RMSE of the spatial-frequency estimates in radians:
\begin{equation}
    \mathrm{RMSE}
    \triangleq
    \sqrt{
      \mathbb{E}\!\left\{
        \frac{1}{d}\sum_{k=1}^{d} (\widehat{\mu}_k - \mu_{\text{true},k})^2
      \right\}
    },
    \label{eq:RMSEdef}
\end{equation}
estimated from $R{=}10^{4}$ independent trials (see Table~\ref{tab:sim_params}). The stochastic Cram\'er--Rao bound~\cite{vanTrees}:
\begin{equation}
    \mathrm{CRB}(\boldsymbol{\mu})
    \triangleq
    \frac{1}{d}\operatorname{tr}\!\big(\mathbf{J}(\boldsymbol{\mu})^{-1}\big),
    \label{eq:CRBdef}
\end{equation}
plotted as $\sqrt{\mathrm{CRB}(\boldsymbol{\mu})}$ with the gap-to-CRB in dB defined as $10\log_{10}\!\big(\mathrm{RMSE}/\sqrt{\mathrm{CRB}}\big)$. The failure rate, defined relative to a CRB-based threshold:
\begin{equation}
    P_{\mathrm{fail}}
    \triangleq
    \mathbb{P}\Big(
      \max_{k} |\widehat{\mu}_k - \mu_{\text{true},k}|
      >
      k_{\mathrm{thr}}\sqrt{\mathrm{CRB}(\boldsymbol{\mu})}
    \Big),
    \label{eq:failureRatedef}
\end{equation}
with threshold factor $k_{\mathrm{thr}}{=}3$. In the plots we estimate $P_{\mathrm{fail}}$ via the empirical fraction of failures across trials and report Wilson 95\% confidence intervals per ASNR. Distributional diagnostics that complement RMSE and failure-rate curves are reported in Appendix~\ref{app:diagnostics}, Sec.~\ref{subsec:ECDFs}--\ref{subsec:LPA}.

\subsection{Main results}

\subsubsection{RMSE and gap-to-CRB}
\begin{figure*}[!t]
  \centering
  \subfloat[RMSE vs. ASNR.]{
    \includegraphics[width=0.49\textwidth]{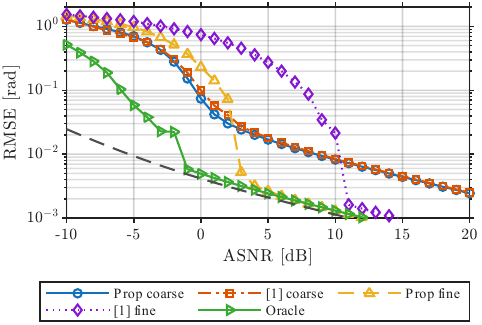}%
    \label{fig:rmse_vs_asnr}
  }\hfill
  \subfloat[Gap-to-CRB vs. ASNR.]{
    \includegraphics[width=0.48\textwidth]{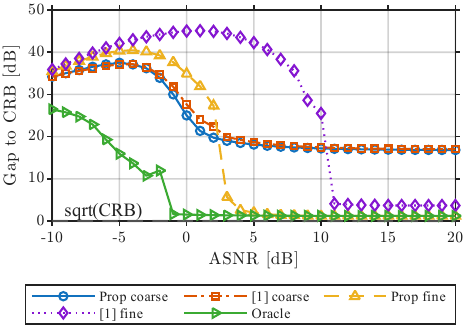}%
    \label{fig:gap_to_crb_vs_asnr}
  }
  \caption{Angle estimation performance versus ASNR for
  $N_{\mathrm{RF}}^{\mathrm{coarse}}=12 \rightarrow N_{\mathrm{RF}}^{\mathrm{fine}}=6$:
  (a)~RMSE; (b)~gap to the CRB.
  The fine-stage RMSE in~(a) temporarily exceeds the coarse baselines at low ASNR
  (below approximately $2$~dB); this crossover occurs when sector-assignment
  reliability is insufficient for accurate covariance-guided beam selection
  (see Section~\ref{sec:num_results}).}
  \label{fig:rmse_and_gap_subfig}
\end{figure*}

Fig.~\ref{fig:rmse_and_gap_subfig} shows RMSE and gap-to-CRB versus ASNR for
$N_{\mathrm{RF}}^{\mathrm{coarse}}{=}12$ and $N_{\mathrm{RF}}^{\mathrm{fine}}{=}6$.
The two coarse-only baselines (centro-symmetric mask vs.\ non-symmetric mask) exhibit
similar trends and remain several dB away from the CRB across the SNR range, as expected
from the reduced aperture.
By contrast, the proposed covariance-guided fine stage yields a pronounced improvement:
the RMSE curve closely tracks that of the oracle beam configuration, and the gap-to-CRB
remains within approximately $1$--$2$~dB for $\mathrm{ASNR}\ge 4$~dB.
The sectorization-based fine stage from \cite{TensorESPRIT_Beamspace} also improves over
the coarse estimates, but requires roughly $4$--$6$~dB higher ASNR to reach a comparable
near-CRB regime. In the considered $M{=}32$, $d{=}3$ scenario, the proposed beam selection
therefore extracts more information per active beam under the same RF-chain
budget.

A transition region is nonetheless visible at low ASNR, where the proposed fine-stage
RMSE can temporarily exceed both coarse-baseline curves.
This behavior is a direct consequence of the coarse-to-fine design: when the ASNR is
insufficient for reliable sector assignment and covariance reconstruction, the fine stage
commits to a narrow beam subset guided by an inaccurate localization cue.
Concentrating the RF-chain budget into a beam subset that excludes part of the source
energy then degrades the signal-to-noise ratio at the ESPRIT input and increases the
effective condition number of the beamspace observation matrix.
The fine stage is therefore best characterized as a reliability-dependent refinement:
it yields a consistent gain over the coarse stage whenever the covariance-capture score
can be trusted, and incurs a transient penalty below the reliability threshold.
The failure-rate curves in Fig.~\ref{fig:failureRates} and the sector-edge analysis in
Section~\ref{subsec:ablations} quantify this threshold more precisely.

\subsubsection{Reliability and threshold behaviour}
\begin{figure}[!t]
\centering
\includegraphics[width=1\columnwidth]{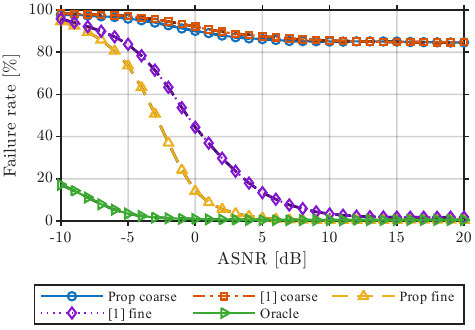}
\caption{Failure rates vs. ASNR.}
\label{fig:failureRates}
\end{figure}

Fig.~\ref{fig:failureRates} reports failure probabilities as a function of ASNR using
the CRB-based threshold in \eqref{eq:failureRatedef}.
The proposed fine stage achieves less than $10\%$ failure probability already around
$0$--$1$~dB, whereas the fine stage of \cite{TensorESPRIT_Beamspace} requires
approximately $5$--$6$~dB to reach a similar reliability level.
At high ASNR both fine-stage pipelines eventually approach zero failures, but the
covariance-guided selector enters the reliable regime much earlier.
This behaviour is consistent with the smaller performance gap to the oracle in
Fig.~\ref{fig:rmse_and_gap_subfig}.
The reliability threshold visible here corresponds to the ASNR below which
sector-assignment errors become frequent; this regime is also where the transient
RMSE crossover in Fig.~\ref{fig:rmse_vs_asnr} occurs, as discussed above.

\subsection{Ablations and stress tests}
\label{subsec:ablations}

To better understand the behavior of the proposed framework, we next perform a set of ablation and stress-test experiments. We systematically vary the number of RF chains allocated to the coarse and fine stages, the total beam budget, the number of snapshots, and the position of the dominant path relative to sector boundaries. These experiments complement recent deep-learning-based studies on mmWave channel estimation and hybrid beamforming under changing network
conditions~\cite{Zhang2025_RRNet_HSR, Guo2024_DL_CE_FB} by providing a detailed picture of how a model-based covariance-guided ESPRIT framework behaves when hardware and training resources are reallocated. In particular, the Pareto figure in Fig.~\ref{fig:A1_pareto} of Appendix~\ref{app:ablations} confirms that the default $12\to 6$ RF-chain allocation lies near the knee of the accuracy--runtime
trade-off for the considered $M{=}32$, $d{=}3$ scenario, providing empirical support for the design choice discussed in Section~\ref{sec:num_results}.

To isolate the sources of performance gains, we include three ablations: (i) accuracy--runtime trade-off under a dynamic RF (beam) budget, (ii) sector-edge stress test with two sources, and (iii) fixed fine-stage beam budget $K_g{=}K_f$ comparing covariance-guided selection with pure sectorization.

\subsubsection{Sector-edge stress test with two sources}
\begin{figure*}[!t]
  \centering
  \subfloat[RMSE vs. $\delta/(w_{\mathrm{sec}}/2)$.]{
    \includegraphics[width=0.48\textwidth]{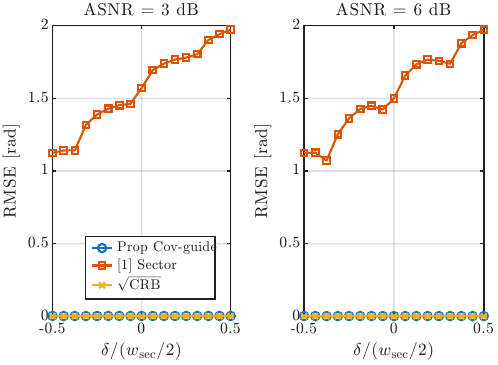}%
    \label{fig:A2_RMSE}
  }\hfill
  \subfloat[Failure rate vs. $\delta/(w_{\mathrm{sec}}/2)$.]{
    \includegraphics[width=0.48\textwidth]{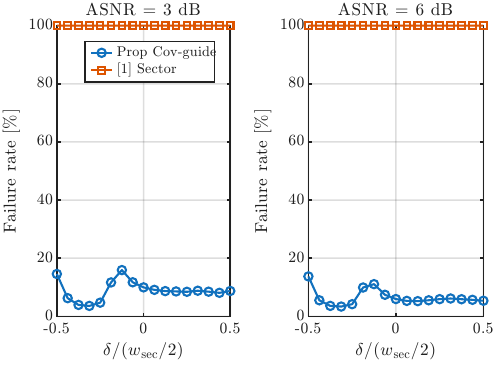}%
    \label{fig:A2_FailRate}
  }
  \caption{Sector-edge stress test with two sources versus normalized displacement $\delta/(w_{\mathrm{sec}}/2)$: (a) RMSE; (b) failure rate.}
  \label{fig:A2_sector_edge}
\end{figure*}

In this ablation we stress-test robustness near sector boundaries. We consider $d{=}2$ uncorrelated sources with power vector
\[
\mathbf{p} = [p_1,p_2]^{\mathsf{T}} = [0.95,\,0.5]^{\mathsf{T}},
\quad
\mathbf{R}_{\mathbf{s}} = \operatorname{diag}(\mathbf{p}),
\]
and spatial frequencies $\boldsymbol{\mu} = [\mu_1,\mu_2]^{\mathsf{T}}$. The first source is fixed at $\mu_1=-2.1$, while the second source is swept across the boundary between two adjacent DFT beams. Let $\kappa_{\mathrm{bdy}}$ and $\kappa_{\mathrm{bdy}}{+}1$ denote the corresponding DFT beam indices with spatial frequencies $\gamma_{\kappa_{\mathrm{bdy}}}$ and $\gamma_{\kappa_{\mathrm{bdy}}+1}$. We define the boundary spatial frequency
\[
\mu_{\mathrm{edge}}
\triangleq
\tfrac{1}{2}\big(\gamma_{\kappa_{\mathrm{bdy}}} + \gamma_{\kappa_{\mathrm{bdy}}+1}\big),
\]
and parameterize the displacement of the second source by $\delta \triangleq \mu_2 - \mu_{\mathrm{edge}}$. The horizontal axis in Figs.~\ref{fig:A2_RMSE} and \ref{fig:A2_FailRate} is normalized as $\delta/(w_{\mathrm{sec}}/2)$, where $w_{\mathrm{sec}}$ is the sector width.

The coarse stage again uses $N_{\mathrm{RF}}^{\mathrm{coarse}}{=}12$ RF chains and FBA-TLS-ESPRIT on the centro-symmetric mask. For the fine stage we compare the proposed covariance-guided selector with the sectorization baseline at beam budget $K_f{=}2$, both followed by beamspace Unitary ESPRIT.

Figure~\ref{fig:A2_RMSE} plots the RMSE of the spatial-frequency estimates versus the normalized boundary offset at $\mathrm{ASNR} = 3$ and $6$~dB. The sectorization baseline exhibits a broad region of elevated error around the boundary and, in the low-SNR case, does not approach the CRB for any offset. The covariance-guided beams track the CRB more closely over a wide range of offsets and reduce the peak error in the immediate vicinity of the boundary. As ASNR increases, the gap between the methods shrinks, but the covariance-guided selector remains more robust to small displacements around the sector edges in this stress test.

The corresponding failure probabilities in Fig.~\ref{fig:A2_FailRate} show a consistent trend. At low ASNR, the sectorization-based fine stage experiences failure rates close to $100\%$ for a wide range of offsets, whereas the proposed method keeps the failure rate low except in a narrow neighbourhood of the exact boundary. At higher ASNR, both methods see reduced failure rates, but the covariance-guided selector still exhibits lower failure probability across most offsets.
These results suggest that, once the fine beams are selected in a data-aware manner, the residual sector-edge sensitivity is mainly due to the coarse-sector definition rather than an intrinsic limitation of ESPRIT in beamspace. The transient low-ASNR penalty identified in Fig.~\ref{fig:rmse_vs_asnr}(a) is consistent with this picture: at the ASNR values where sector-edge failures are elevated, covariance-guided selection cannot fully compensate for an unreliable coarse localization cue, whereas at moderate-to-high ASNR the data-aware beam assignment recovers the near-CRB regime across the full offset range.

Additional distributional diagnostics are provided in Appendix~\ref{app:diagnostics}, and additional ablations are provided in Appendix~\ref{app:ablations}.

\subsection{Phase-shifter quantization robustness}
\label{subsec:phase_quant}
\normalcolor

\begin{figure}[!t]
\centering
\includegraphics[width=1\columnwidth]{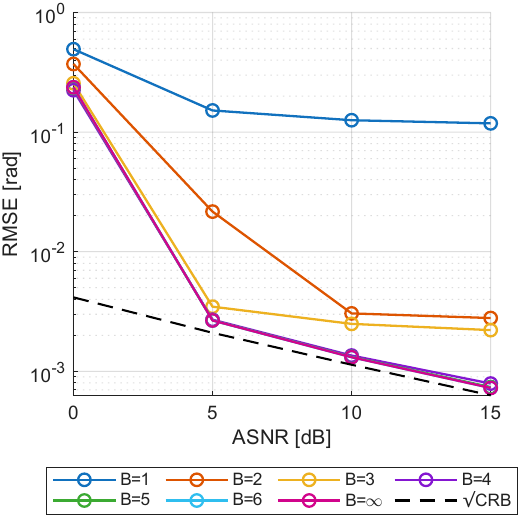}
\caption{Phase-shifter quantization robustness ($M{=}32$, $d{=}3$,
  $N{=}100$ snapshots, $R{=}10^4$ trials). RMSE versus ASNR for
  $B$-bit phase-quantized DFT dictionaries ($B \in \{1,2,3,4,5,6\}$,
  dashed lines for $B{=}1,2$) and the ideal continuous-phase reference
  ($B{=}\infty$, solid). Dotted line: $\sqrt{\mathrm{CRB}}$.
  For $B \ge 4$ the RMSE is within $9\%$ of the $B{=}\infty$ baseline
  at $15$~dB; all $B \ge 4$ curves converge at $0$~dB, confirming that
  the low-ASNR bottleneck is coarse-stage resolution, not fine-stage
  quantization.}
\label{fig:phase_quant}
\end{figure}

Section~\ref{subsec:HBA} shows that $B$-bit phase quantization displaces each DFT column by at most $2\sin(\pi/2^{B+1})$ in $\ell_2$-norm, a hardware-distortion bound that decreases rapidly with $B$ and reaches $0.20$ at $B{=}4$. To quantify the end-to-end impact of this dictionary distortion on DoA estimation accuracy, we perform a dedicated Monte Carlo experiment ($R{=}10^4$ trials) matching the setup of Section~\ref{subsec:exp_setup}: $M{=}32$, $d{=}3$, $N{=}100$ snapshots, and the three-path power vector of the main experiment. The coarse stage uses element-space subsampling and is $B$-invariant; only the fine-stage analog combiner is replaced by a $B$-bit quantized DFT dictionary. Figure~\ref{fig:phase_quant} plots RMSE versus ASNR for $B \in \{1,2,3,4,5,6,\infty\}$.

For $B \ge 4$ the RMSE curves are essentially indistinguishable from the ideal $B{=}\infty$ baseline: at $15$~dB the overhead is $8.8\%$ in RMSE for $B{=}4$ and below $2\%$ for $B \ge 5$, confirming graceful degradation. The values $B{=}1$ and $B{=}2$ fall below any practical operating threshold: failure rates reach $100\%$ and ${\sim}64\%$, respectively, at $15$~dB ASNR, confirming that fewer than 3 bits are inadequate for fine-stage beamforming regardless of SNR. At $0$~dB ASNR all curves for $B \ge 4$ converge to the same RMSE as the ideal-DFT reference (spread ${<}6\%$), with failure rates of $0.141$--$0.143$, indicating that the performance bottleneck at low ASNR is coarse-stage angular resolution rather than fine-stage phase quantization. The pipeline-fidelity check ($\mathrm{RMSE}(B{=}\infty,\,15~\mathrm{dB})= 7.26{\times}10^{-4}$~rad) matches the main-experiment benchmark to within $0.3\%$.

\section{Conclusion}
\label{sec:conclusion}
\normalcolor
This paper has proposed a covariance-guided DFT beam selection and beamspace ESPRIT framework for hybrid analog/digital mmWave sensor front-ends. By combining a virtual fully digital subarray, Toeplitz--PSD covariance fitting, analytically motivated DFT beam selection, and sparse beamspace Unitary ESPRIT, the framework achieves high-resolution spatial-frequency estimation under strict RF-chain and beam-budget constraints for the considered 32-element ULA with three uncorrelated paths. The core idea is to dedicate a short hybrid training phase to reconstruct a denoised full-aperture covariance matrix and to use this covariance to select contiguous DFT beam subsets that preserve effective aperture while concentrating power around the dominant paths.

Monte Carlo simulations for a 32-element ULA with three paths show that, in the considered scenarios, the proposed framework can reduce the gap to the Cram\'er--Rao bound relative to a sectorization-based baseline constructed from the same DFT codebook and ESPRIT estimator. The covariance-guided method also tends to lower the failure rate and offers competitive accuracy--runtime trade-offs under varying ASNRs, numbers of snapshots, beam budgets, and sector-edge conditions. Overall, the results indicate that covariance-guided beam selection is often more informative than purely energy-based sectorization for a given beam budget, and that a single low-dimensional ESPRIT call, driven by carefully selected beams, can approach the oracle performance in many of the tested operating points. These properties are directly relevant for mmWave sensing applications that are often reported in IEEE Sensors Journal, such as radar-based localization, human-activity sensing, and MIMO-SAR imaging with limited RF budgets~\cite{Su2023JSEN, Rai2021JSEN}. More broadly, the covariance-guided selection principle could be extended to scenarios where a limited RF budget must be shared across multiple sensing
directions or target types, such as joint radar-communication front-ends operating under a strict beam budget~\cite{Leyva2024JSEN}.

The present study uses the unitary $M$-point DFT dictionary and an idealized virtual-subarray mapping; under these assumptions the beam-set score reduces to a covariance-guided energy-window search (Remark~\ref{rem:unitary_dft}), while non-unitary codebooks and hardware impairments make the conditioning terms active and are natural extensions.

The present study is deliberately focused in scope. The numerical validation is restricted to a 32-element ULA, three uncorrelated paths, spatially white receiver noise, and an idealized hybrid front end with a unitary DFT codebook, unless stated otherwise. Within this scope, the main contribution is a covariance-denoised, beam-budgeted refinement strategy for beamspace ESPRIT that consistently reduces the gap to the CRB relative to sectorization above the coarse-stage reliability threshold. Broader claims for coherent multipath, wideband channels, non-DFT codebooks, array mismatch, or near-field operation require additional evidence beyond what is
provided here.

From a sensing-hardware perspective, several further limitations are worth
highlighting. First, we restrict attention to uncorrelated, noncoherent sources and spatially white receiver noise, and we rely on a Toeplitz--PSD approximation of the virtual full-aperture covariance. Coherent or strongly correlated multipath, which frequently arises in specular or cluttered sensing environments, can degrade both the covariance-fitting stage and the ESPRIT estimator and will require additional regularization or element-space preprocessing. Second, the virtual subarray and DFT beams are assumed to be perfectly realized by the hybrid hardware: we ignore mutual coupling, RF-chain gain and phase mismatches, finite-resolution phase shifters, and calibration errors. For finite-resolution phase shifters specifically, the effective beam dictionary departs from the ideal unitary DFT case (see Section~\ref{subsec:HBA}), activating the conditioning terms in \eqref{eq:preservationScore}--\eqref{eq:score_def}; the general score formulation is architecturally designed to handle this regime, and a systematic characterization of the graceful-degradation boundary under $B$-bit phase quantization---including the interaction with covariance-fit quality at finite snapshot counts---is an important direction for future work. A more comprehensive robustness study covering mutual coupling, gain/phase calibration errors, and experimental validation with over-the-air or hardware-in-the-loop measurements remains open for future work toward deployable mmWave sensing front-ends.

Several additional directions remain open for future research. A natural extension is to integrate the proposed covariance-guided beamspace ESPRIT framework with wideband OFDM channel models and near-field or extremely large-scale MIMO architectures; for scenarios where a larger beam budget is available, multi-dimensional tensor ESPRIT formulations \cite{ZhangCheng2024TWC,TensorESPRIT_Beamspace} could further improve estimation accuracy by exploiting Kronecker separability across spatial and delay--Doppler dimensions. Another direction is to incorporate parametric or learned priors on the covariance structure, potentially combining the present approach with deep-learning-based channel estimation or hybrid beamforming. Finally, it would be interesting to explore adaptive beam-budget and RF-chain allocation policies that use intermediate estimation quality indicators to reassign RF resources across angular sectors, subcarriers, and probing intervals in dynamic mmWave sensing scenarios.



\appendices

\setcounter{figure}{0}
\setcounter{table}{0}
\setcounter{equation}{0}
\setcounter{algorithm}{0}
\renewcommand{\thefigure}{S\arabic{figure}}
\renewcommand{\thetable}{S\arabic{table}}
\renewcommand{\theequation}{S\arabic{equation}}
\renewcommand{\thealgorithm}{S\arabic{algorithm}}

\section{Additional Diagnostics for the Main ASNR Study}
\label{app:diagnostics}
In this section we provide distributional and subspace-alignment diagnostics that complement the RMSE, gap-to-CRB, and failure-rate curves reported in the main manuscript. Unless stated otherwise, we use the same simulation parameters as Table~I.

\subsection{Distributional Shape via ECDFs}
\label{subsec:ECDFs}
\begin{figure}[!t]
\centering
\includegraphics[width=0.8\columnwidth]{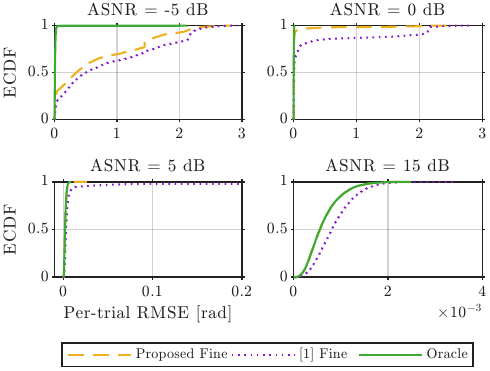}
\caption{ECDFs of per-trial estimation error at ASNR $=-5, 0, 5, 15$~dB.}
\label{fig:ECDFs}
\end{figure}

Fig.~\ref{fig:ECDFs} reports empirical CDFs (ECDFs) of the per-trial estimation error at representative ASNR values. The ECDFs help interpret the failure-rate transition and the tail behavior that is not visible from RMSE alone.

\subsection{Subspace Alignment via Largest Principal Angle}
\label{subsec:LPA}
\begin{figure}[!t]
\centering
\includegraphics[width=0.7\columnwidth]{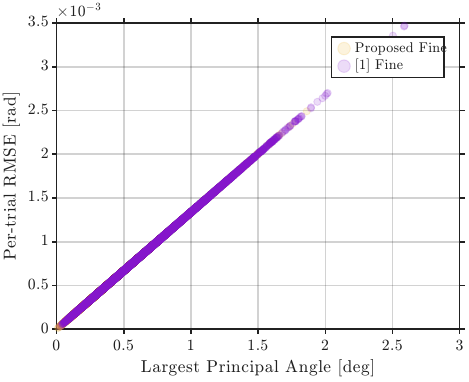}
\caption{LPA--error scatter at ASNR $=15$~dB.}
\label{fig:LPA}
\end{figure}

Fig.~\ref{fig:LPA} relates estimation error to signal-subspace alignment at $\mathrm{ASNR}=15$~dB. Let $\mathbf{U}_{\mathrm{S}}$ and $\widehat{\mathbf{U}}_{\mathrm{S}}$ contain orthonormal bases for the true and estimated $d$-dimensional signal subspaces, respectively. We define the largest principal angle (LPA) as
\begin{equation}
\mathrm{LPA}
\triangleq
\frac{180}{\pi}
\cos^{-1}\!\left(
  \sigma_{\min}\!\big(
    \mathbf{U}_{\mathrm{S}}^{\mathsf{H}}\widehat{\mathbf{U}}_{\mathrm{S}}
  \big)
\right),
\label{eq:LPAdef_supp}
\end{equation}
where $\sigma_{\min}(\cdot)$ denotes the smallest singular value. Smaller LPA indicates tighter subspace alignment.

\section{Additional Ablations and Stress Tests}
\label{app:ablations}
In this section we collect ablations that clarify accuracy--runtime behavior and fine-stage beam-budget effects.

\subsection{Accuracy--Runtime Trade-off Under a Dynamic Fine-Stage Beam Budget}
We compare four fine-stage configurations, keeping the coarse stage fixed:
\[
\text{Cov~}12{\rightarrow}6,\quad
\text{Cov~}12{\rightarrow}12,\quad
\text{Sect~}12{\rightarrow}6,\quad
\text{Sect~}12{\rightarrow}12,
\]
where ``Cov'' denotes covariance-guided selection from the fitted Toeplitz--PSD covariance and ``Sect'' denotes the sectorization-based allocation baseline from \cite{TensorESPRIT_Beamspace}. Table~\ref{tab:A1_timing} reports median runtime breakdowns and DOA RMSE at representative ASNR values.

\begin{table*}[!t]
    \centering
    \small
    \caption{Median runtime and DOA RMSE for the dynamic fine-stage budget ablation. Timings are medians over $10^{4}$ Monte Carlo trials.}
    \label{tab:A1_timing}
    \begin{tabular}{l c c c c c c}
        \toprule
        Configuration & ASNR [dB] & RMSE [rad] & $t_{\mathrm{cov}}$ [ms] & $t_{\mathrm{sel}}$ [ms] & $t_{\mathrm{ES}}$ [ms] & $t_{\mathrm{total}}$ [ms] \\
        \midrule
        Cov~12$\rightarrow$6  & 3 & 0.015512  & 2.0945 & 0.86825 & 0.36675 & 3.3961 \\
        Cov~12$\rightarrow$6  & 6 & 0.0024718 & 2.0889 & 0.86995 & 0.36560 & 3.4057 \\
        Cov~12$\rightarrow$12 & 3 & 0.028267  & 2.0982 & 0.73865 & 0.42450 & 3.3275 \\
        Cov~12$\rightarrow$12 & 6 & 0.0063330 & 2.1107 & 0.73895 & 0.42245 & 3.3426 \\
        Sect~12$\rightarrow$6 & 3 & 0.45542   & 1.8838 & 0.26605 & 0.23610 & 2.4263 \\
        Sect~12$\rightarrow$6 & 6 & 0.19915   & 1.8861 & 0.26685 & 0.23605 & 2.4278 \\
        Sect~12$\rightarrow$12& 3 & 0.014777  & 1.9030 & 0.27000 & 0.29430 & 2.5057 \\
        Sect~12$\rightarrow$12& 6 & 0.0063796 & 1.9008 & 0.27055 & 0.29010 & 2.5061 \\
        \bottomrule
    \end{tabular}
\end{table*}

\begin{figure}[!t]
    \centering
    \includegraphics[width=0.4\textwidth]{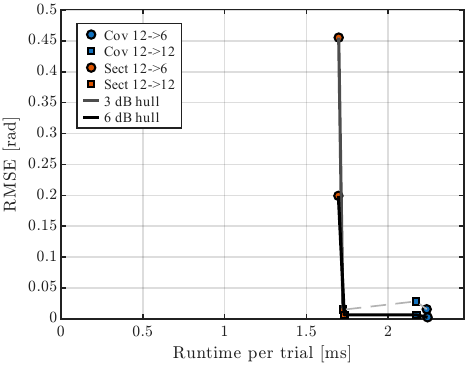}
    \caption{Pareto trade-off between median runtime per Monte Carlo trial and DOA RMSE for the four fine-stage configurations.}
    \label{fig:A1_pareto}
\end{figure}

Fig.~\ref{fig:A1_pareto} summarizes the empirical Pareto trade-off between median runtime per trial and DOA RMSE for the four fine-stage configurations.

\subsection{Fixed Per-Sector Beam Budget $K_g$: Covariance-Guided Selection vs.\ Sectorization}
We fix $K_g\in\{2,3,4\}$ and compare it with pure sectorization as in \cite{TensorESPRIT_Beamspace}, using the same fine-stage ESPRIT estimator.

\begin{figure}[!t]
\centering
\includegraphics[width=0.8\columnwidth]{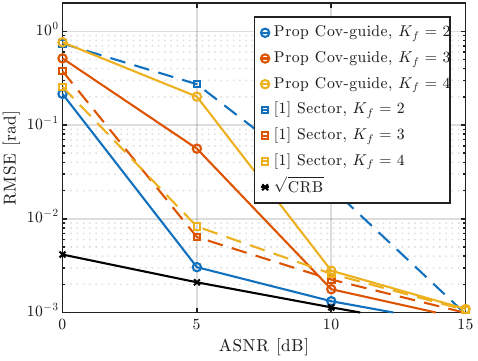}
\caption{RMSE vs.\ ASNR for $K_g \in \{2,3,4\}$.}
\label{fig:S1_RMSE}
\end{figure}

\begin{figure}[!t]
\centering
\includegraphics[width=0.8\columnwidth]{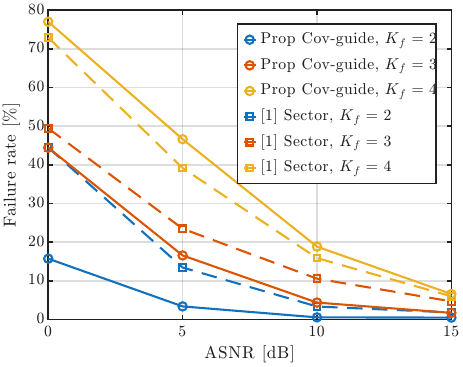}
\caption{Failure rates vs.\ ASNR for $K_g \in \{2,3,4\}$.}
\label{fig:S1_Fail}
\end{figure}

\section{Real-Valued Quadratic-Program Formulations}
\label{app:qp_real}

This appendix records real-valued quadratic-program (QP) formulations that implement (i) the NNLS power/noise fitting step and (ii) the Toeplitz--PSD projection step described in the main manuscript.

\subsection{NNLS Power and Noise-Variance Fitting}
Given the coarse spatial-frequency estimates, define the subarray steering matrix $\mathbf{A}_{\mathcal{M}}\in\mathbb{C}^{N_{\mathrm{RF}}^{\mathrm{coarse}}\times d}$ and the sample covariance $\widehat{\mathbf{R}}_{\mathcal{M}}$. The NNLS model can be written in complex form as
\begin{equation}
\widehat{\boldsymbol{\xi}}
=
\arg\min_{\mathbf{z}\in\mathbb{R}_+^{d+1}}
\big\| \mathbf{G}_{\mathcal{M}}\mathbf{z} - \mathrm{vec}(\widehat{\mathbf{R}}_{\mathcal{M}}) \big\|_2^2,
\label{eq:nnls_covfit_supp}
\end{equation}
where $\mathbf{G}_{\mathcal{M}}$ matches the definition in the main manuscript and
$\boldsymbol{\xi}=[p_1,\ldots,p_d,N_0]^{\mathsf{T}}$.

For implementation, we convert \eqref{eq:nnls_covfit_supp} into a real QP by stacking real/imaginary parts. Let
\[
\mathbf{y} \triangleq
\begin{bmatrix}
\Re\{\mathrm{vec}(\widehat{\mathbf{R}}_{\mathcal{M}})\}\\
\Im\{\mathrm{vec}(\widehat{\mathbf{R}}_{\mathcal{M}})\}
\end{bmatrix},
\qquad
\mathbf{C} \triangleq
\begin{bmatrix}
\Re\{\mathbf{G}_{\mathcal{M}}\}\\
\Im\{\mathbf{G}_{\mathcal{M}}\}
\end{bmatrix}.
\]
Then \eqref{eq:nnls_covfit_supp} is equivalent to
\begin{equation}
\min_{\mathbf{x}\ge \mathbf{0}}
\frac{1}{2}\,\mathbf{x}^{\mathsf{T}}\mathbf{H}_{\mathrm{pow}}\mathbf{x}
- \mathbf{q}_{\mathrm{pow}}^{\mathsf{T}}\mathbf{x},
\label{eq:qp_power_supp}
\end{equation}
with $\mathbf{H}_{\mathrm{pow}} = 2\mathbf{C}^{\mathsf{T}}\mathbf{C}$ and $\mathbf{q}_{\mathrm{pow}}=2\mathbf{C}^{\mathsf{T}}\mathbf{y}$. In practice we add a small ridge $\varepsilon\mathbf{I}$ to $\mathbf{H}_{\mathrm{pow}}$ for numerical stability.

\subsection{Toeplitz--PSD Projection via PSD-Sampling Constraints}
Let $\widehat{\mathbf{R}}_{\mathrm{full}}$ denote the preliminary full-aperture covariance formed from the fitted powers and noise variance. We compute the Toeplitz--PSD projected covariance as
\begin{IEEEeqnarray}{rCl}
\widehat{\mathbf{R}}_{\mathrm{T}}
&=&
\arg\min_{\mathbf{T}}
\big\|\mathbf{T}-\widehat{\mathbf{R}}_{\mathrm{full}}\big\|_F^2
\nonumber \\
&\text{s.t.}& \,
\mathbf{T}=\mathbf{T}^{\mathsf{H}},\ \mathbf{T}\ \text{Toeplitz},\ \mathbf{T}\succeq \mathbf{0}.
\label{eq:toeplitz_psd_projection_supp}
\end{IEEEeqnarray}
A Hermitian Toeplitz matrix is parameterized by its first column $\mathbf{t}=[t_0,t_1,\ldots,t_{M-1}]^{\mathsf{T}}$ with $t_0\in\mathbb{R}$. We stack real parameters into
\[
\mathbf{z}\triangleq
\begin{bmatrix}
t_0,\ \Re\{t_1\},\ldots,\Re\{t_{M-1}\},\ \Im\{t_1\},\ldots,\Im\{t_{M-1}\}
\end{bmatrix}^{\mathsf{T}}
\]
where $\mathbf{z}\in\mathbb{R}^{2M-1}$. Then the Frobenius objective becomes a real least-squares objective in $\mathbf{z}$, yielding a QP of the form
\begin{equation}
\min_{\mathbf{z}}
\frac{1}{2}\,\mathbf{z}^{\mathsf{T}}\mathbf{H}_{\mathrm{toe}}\mathbf{z}
- \mathbf{q}_{\mathrm{toe}}^{\mathsf{T}}\mathbf{z},
\label{eq:qp_toeplitz_supp}
\end{equation}
for suitable $\mathbf{H}_{\mathrm{toe}},\mathbf{q}_{\mathrm{toe}}$.

To enforce $\mathbf{T}\succeq \mathbf{0}$, we impose nonnegativity of the sampled power spectral density on a dense grid $\{\omega_m\}_{m=1}^{M_{\mathrm{grid}}}$ over $[-\pi,\pi)$:
\[
\lambda(\omega_m)\ge 0,\qquad m=1,\ldots,M_{\mathrm{grid}},
\]
which yields linear inequalities $\mathbf{A}_{\mathrm{psd}}\mathbf{z}\ge \mathbf{0}$. The implemented Toeplitz--PSD projection is therefore
\begin{equation}
\min_{\mathbf{z}}
\frac{1}{2}\,\mathbf{z}^{\mathsf{T}}\mathbf{H}_{\mathrm{toe}}\mathbf{z}
- \mathbf{q}_{\mathrm{toe}}^{\mathsf{T}}\mathbf{z}
\quad\text{s.t.}\quad
\mathbf{A}_{\mathrm{psd}}\mathbf{z}\ge \mathbf{0}.
\label{eq:qp_toeplitz_psd_supp}
\end{equation}
After solving \eqref{eq:qp_toeplitz_psd_supp}, we reconstruct $\widehat{\mathbf{R}}_{\mathrm{T}}$ from the optimal Toeplitz parameters.

\section{Covariance-Guided DFT Beam Selection Pseudocode}
\label{app:beam_selection}

This appendix provides pseudocode for the covariance-guided contiguous DFT beam selection described in the main manuscript. The input denoised signal covariance is
\[
\widetilde{\mathbf{R}}_{\mathrm{sig}}
\triangleq
\widehat{\mathbf{R}}_{\mathrm{T}}-\widehat{N}_0\mathbf{I}_M,
\]
consistent with the main text.

\begin{algorithm}[t]
\small
\caption{Covariance-Guided DFT Beam Selection (per sector)}
\label{alg:cov_beam_selection}
\begin{algorithmic}[1]
  \Require Denoised signal covariance $\widetilde{\mathbf{R}}_{\mathrm{sig}}\in\mathbb{C}^{M\times M}$
  \Require Sector-wise beam pools $\{\mathcal{B}_g\}_{g=1}^{G}$, $\mathcal{B}_g\subset\{1,\ldots,M\}$
  \Require Per-sector beam budgets $\{K_g\}_{g=1}^{G}$ with $K_g\ge 2$
  \Require Parameters $\eta\ge 0$, $\alpha\ge 0$, optional pruning parameter $q\in\mathbb{N}$ ($q=0$ disables pruning)
  \Ensure  Contiguous selected beam sets $\mathcal{S}_g^\star\subset\mathcal{B}_g$ for $g=1,\ldots,G$

  \State Form the unitary (FFT-shifted) DFT beamforming matrix $\mathbf{B}=[\mathbf{b}_1,\ldots,\mathbf{b}_M]\in\mathbb{C}^{M\times M}$.
  \State Compute per-beam energies $\rho_m \gets \mathbf{b}_m^{\mathsf{H}}\widetilde{\mathbf{R}}_{\mathrm{sig}}\mathbf{b}_m$ for $m=1,\ldots,M$.

  \For{$g = 1$ \textbf{to} $G$}
    \State Let $\mathcal{B}_g=\{\kappa_1,\ldots,\kappa_{L_g}\}$ be the sorted beam indices in sector $g$.
    \State Clip $K_g$ to satisfy $2 \le K_g \le L_g$.
    \State Initialize candidate window starts: $\mathcal{S}_{\mathrm{start}}\gets\{1,\ldots,L_g-K_g+1\}$.

    \If{$q>0$ \textbf{and} $L_g>K_g$}
      \State Let $\mathcal{C}_{\mathrm{top}}$ be the positions (within $\{1,\ldots,L_g\}$) of the $q$ largest $\rho_{\kappa_\ell}$ in the sector.
      \State Set $h\gets\lceil (K_g-1)/2\rceil$.
      \State Prune: $\mathcal{S}_{\mathrm{start}}\gets\{\,s\in\mathcal{S}_{\mathrm{start}}: s+h\in\mathcal{C}_{\mathrm{top}}\,\}$.
    \EndIf

    \State Initialize $\mathrm{score}^\star\gets-\infty$, $\mathcal{S}_g^\star\gets\emptyset$.

    \ForAll{$s\in\mathcal{S}_{\mathrm{start}}$}
      \State $\mathcal{S}_g(s)\gets\{\kappa_s,\ldots,\kappa_{s+K_g-1}\}$.
      \State $\mathbf{B}_g\gets[\mathbf{b}_m]_{m\in\mathcal{S}_g(s)}\in\mathbb{C}^{M\times K_g}$.
      \State $\mathbf{G}_g\gets\mathbf{B}_g^{\mathsf{H}}\mathbf{B}_g$, \quad
             $\mathbf{C}_g\gets\mathbf{B}_g^{\mathsf{H}}\widetilde{\mathbf{R}}_{\mathrm{sig}}\mathbf{B}_g$.
      \State $\mathrm{cap}\gets\operatorname{tr}\!\big[(\mathbf{G}_g+\eta\mathbf{I})^{-1}\mathbf{C}_g\big]$.
      \State $\kappa_g^2\gets \operatorname{cond}(\mathbf{G}_g)^2$.
      \State $\mathrm{score}\gets \mathrm{cap}/(1+\alpha\,\kappa_g^2)$.
      \If{$\mathrm{score}>\mathrm{score}^\star$}
        \State $\mathrm{score}^\star\gets\mathrm{score}$, \quad $\mathcal{S}_g^\star\gets\mathcal{S}_g(s)$.
      \EndIf
    \EndFor
  \EndFor

  \State \Return $\{\mathcal{S}_g^\star\}_{g=1}^{G}$.
\end{algorithmic}
\end{algorithm}

\end{document}